\documentclass[11pt,a4paper,notoc]{article}
\usepackage{jheppub}
\usepackage{amsmath}         
\usepackage{amssymb}          
\usepackage{bbold}
\usepackage{ulem}
\usepackage{afterpage}
\usepackage{youngtab}
\usepackage{multirow}
\usepackage[capitalise]{cleveref}
\usepackage{slashed}
\usepackage{booktabs}
\usepackage{subcaption}
\usepackage{comment}
\usepackage{rotating}
\usepackage{url}
\usepackage{xcolor}
\usepackage{color}
\usepackage{float}
\definecolor{linkcolor}{rgb}{0,0,0.6}
\definecolor{mygreen}{rgb}{0,0.6,0}
\definecolor{ballblue}{rgb}{0.13, 0.67, 0.8}

\def\red#1{\textcolor{black}{#1}}

\crefname{section}{sec.}{secs.}
\crefname{table}{Tab.}{Tabs.}
\crefname{figure}{Fig.}{Figs.}
\crefname{equation}{Eq.}{Eqs.}
\crefname{appendix}{Appendix}{Appendix}

\usepackage[utf8]{inputenc}

\title{The Cosmological Constant, Dark Matter and the ElectroWeak Scale meet in the Swampland}

\author[a]{Kazem Bitaghsir Fadafan,}%
\emailAdd{bitaghsir@shahroodut.ac.ir}
\affiliation[a]{Faculty of Physics, Shahrood University of Technology, P.O.Box 3619995161 Shahrood, Iran}
\author[b,c,1]{Giacomo~Cacciapaglia\note{Previously at Universite Claude Bernard Lyon 1, CNRS/IN2P3, IP2I UMR 5822, 4 rue Enrico Fermi, F-69100 Villeurbanne, France}}
\affiliation[b]{Laboratoire de Physique Th\'eorique et Hautes \'Energies (LPTHE), UMR 7589, Sorbonne Universit\'e \& CNRS, 4 place Jussieu, 75252 Paris Cedex 05, France}
\affiliation[c]{Quantum Theory Center (QTC) \& D-IAS, Southern Denmark Univ., Campusvej 55, 5230 Odense M, Denmark}
\emailAdd{cacciapa@lpthe.jussieu.fr}

\begin{document}

\abstract{The Swampland program, which looks for low energy theories consistent with quantum gravity, has led to the introduction of a dark dimension stemming from the cosmological constant. We show that the same argument leads to the emergence of the electroweak scale, once the dark dimension is realised in a warped background. A second warped extra dimension at the TeV scale is, therefore, postulated, where the long-standing problem of the hierarchy between the electroweak  and the Planck scales can be addressed. Furthermore, standard model contributions to the cosmological constant are tamed, together with the gravitational ones. In the emergent holistic picture of gravity and gauge interactions, both Planck and the electroweak scales are emergent from a theory with two fundamental scales: $10^{-2}$~eV and $10^{10}$~GeV, which are of geometric origin and, following the Distance Conjecture, natural. Hence, a bridge is established between the two standard models of particle physics and cosmology. }

\maketitle


\section{Introduction}
During the last century, enormous progress in both theory and experiments helped establish solid models that describe the interactions among particles and the cosmological evolution of our Universe. Respectively, they are the standard model (SM) of particle physics \cite{Glashow:1961tr,Weinberg:1967tq,Salam:1968rm}, based on the gauge principle \cite{Yang:1954ek}, and the $\Lambda$CDM model, based on cold dark matter with a cosmological constant $\Lambda$. Both `standard models' entail energy scales that span many orders of magnitude: from the sub-eV scale of the cosmological constant and neutrino masses to the Planck scale ($M_\text{Pl} \sim 10^{28}$~eV), passing through the electroweak scale ($\sim 10^{11}$~eV). Such scales are connected via quantum effects, and the current standard models lack a theoretical understanding of their dynamic relation and relative quantum stability. This issue goes under the name of hierarchy or naturalness problem.

In particular, the cosmological constant \cite{Einstein:1917ce} remains the most mysterious one: we do not have a clear understanding of its origin nor how its value emerges in a quantum context \cite{Weinberg:1988cp}, the latter responsible for making it cosmologically relevant only at present days \cite{Zlatev:1998tr}. Many attempts can be found in the literature to offer a dynamical origin for the cosmological constant, the most famous example being quintessence \cite{Caldwell:1997ii}, a special form of cosmic fluid generated typically by a time-varying scalar field \cite{Ratra:1987rm,Wetterich:1987fm}, although low-redshift data prefer uncoupled quintessence models at odds with the local Hubble constant determinations \cite{Banerjee:2020xcn} (this tension can be avoided in coupled theories, see \cite{Agrawal:2019dlm,DiValentino:2019jae}).
There have been also attempts to relate the value of the cosmological constant to that of neutrino masses, for instance drawing from the weak gravity conjecture \cite{Ibanez:2017kvh}.

Recently, a new proposal has been formulated based on the Swampland program, which aims at relating the consistency of low energy physics models to the ultraviolet embedding in a consistent quantum theory of gravity, see review paper \cite{Palti:2019pca} and references therein. Typically, the consistency is tested via a set of conjectures, which connect different aspects of gravity and quantum field theory. Among these, the Distance Conjecture (DC) \cite{Ooguri:2006in} predicts that infinite towers of states emerge as one approaches infinite distance limits in moduli space. These states become exponentially light, and their presence is expected to lead to the breakdown of the effective field theory. It has been argued that, as a consequence of the DC, the value of the 4-dimensional (4D) cosmological constant $\Lambda$ is associated to the existence of a single large extra dimension with characteristic size of the order of a fraction of eV \cite{Lust:2019zwm}. This construction has been baptised Dark Dimension (DD) \cite{Montero:2022prj} as only gravity can probe it, while the SM fields must remain localised on a 4D manifold. The main driver is a relation between the value of the effective 4D cosmological constant and the mass gap $m$ in the towers of states stemming from the extra dimension:
\begin{equation} \label{eq:DDconj}
    M_\text{Pl}^{\alpha-4} \Lambda \sim (\lambda m)^\alpha\,,
\end{equation}
where $\alpha = 4$ and $\lambda$  is a generic coupling of order $10^{-1} \div 10^{-3}$.
An explicit computation leading to Eq.\eqref{eq:DDconj} is not possible, even though the emergence of light states in the limit of $\Lambda \to 0$ has been verified in all string theory vacua with a well-understood higher-dimensional solution \cite{Lust:2019zwm}. A computation in effective low energy quantum field theory is complicated by the handling of divergences. In fact, this relation has been questioned by arguments based on the existence of a natural cut-off in extra dimensional theories \cite{Branchina:2023ogv,Branchina:2023rgi}, however, more appropriate handling of the cut-off seems to confirm the result in Eq.~\eqref{eq:DDconj} \cite{Burgess:2023pnk,Anchordoqui:2023laz}. Swampland arguments also predict the existence of a new species scale \cite{Dvali:2007hz, Montero:2022prj}
\begin{equation}
    \hat{M} \sim \hat{\lambda}^{-1/3} \Lambda^{1/12} M_\text{Pl}^{2/3} \sim 10^{9} \div 10^{10}~\text{GeV}\,,
\end{equation}
where $\hat{\lambda}$ is another coupling constant, in principle different from $\lambda$ in Eq.~\eqref{eq:DDconj}. This new scale can be associated with the fundamental Planck scale in the ultraviolet completion of the DD. 
Phenomenologically, the DD proposal offers a candidate for dark matter in the form of the tower of gravitational states \cite{Gonzalo:2022jac,Obied:2023clp} as well as a connection with the generation of neutrino masses \cite{Montero:2022prj,Anchordoqui:2023wkm}. Primordial black holes provide an additional component for dark matter \cite{Anchordoqui:2022tgp,Anchordoqui:2022txe} due to the non-standard cosmology of the DD \cite{Anchordoqui:2022svl}. Finally, the realisation of the DD as a strongly warped throat has been studied \cite{Blumenhagen:2022zzw}, as well as a Swampland-motivated connection to the supersymmetry breaking scale at around the TeV scale \cite{Anchordoqui:2023oqm}.

It should be noted that, while the DD proposal tames the quantum contribution of gravity to the cosmological constant via Eq.~\eqref{eq:DDconj}, the contribution of the Standard Model fields remains unaddressed. Being localised on a 4D manifold, they give a localised contribution of the order of the electroweak scale to the fourth power. Hence, it would be crucial to include the electroweak scale in this program and relate it to the measured value of the cosmological constant scale. This is the main goal of our work.

\vspace{0.5cm}

In this paper, we build upon the DD proposal by applying the same Swampland-inspired conjecture above the eV scale. In fact, at energies larger than the scale of the DD, space-time is effectively 5-dimensional (5D) and characterised by its own 5D cosmological constant. Once the DD is realised as a warped dimension, this observation leads naturally to the emergence of the electroweak scale, supported by the existence of another large extra dimension at the TeV scale. Such new dimension can be probed by all SM fields and dynamically explain the value of the Higgs vacuum expectation value, and providing possible signatures at current and future colliders. This proposal solves the hierarchy problem between the electroweak scale and the fundamental Planck scale $\hat{M}$ by connecting both to the cosmological constant (see \cite{Referee2Castellano:2023qhp} for an alternative proposal along the same lines). In our construction, therefore, there will be only two geometrical parameters, the scale of the DD at the meV, and the fundamental Planck scale, $\hat{M} \sim 10^{10}$~GeV. As long as the extra dimensional space is stabilised, the two values are natural at quantum level.

\section{Towards a natural dark dimension}

To properly study a warped DD, it would be necessary to derive the warped geometry from a consistent theory of gravity, such as string theory. As the Swampland program aims at imposing quantum gravity constraints on low energy effective field theories, it is crucial to ensure that the chosen warped background has an ultraviolet completion. Some studies of warped geometries have been conducted within string theory. However, in this paper we propose to initially investigate the main parameters within a Randall-Sundrum background \cite{Randall:1999ee}. Subsequently, we will extend our analysis to examples of warped geometries originating from string theory.

\subsection{A warped dark dimension}

In our scenario, therefore, the DD consists of a warped throat {\it {\`a} la} Randall--Sundrum \cite{Randall:1999ee}, characterised by the following 5D metric
\begin{equation} \label{eq:metric5}
    ds^2 = e^{- 2 k r_c |\phi|} \eta_{\mu \nu} dx^\mu dx^\nu + r_c^2 d\phi^2\,,
\end{equation}
where $\phi \in [0,\pi]$ spans the extra space-like coordinate, $k$ is the curvature, $r_c$ is the radius of the extra dimension.
The 5D theory is characterised by an effective 5D Planck mass $M_5$, a bulk cosmological constant $\Lambda_5$, and the mass gap of the tower of states $m$, which are related to the fundamental parameters in the metric and to the 4D Planck scale $M_\text{Pl} = 1.22\cdot 10^{19}$~GeV. We recall that, to be a solution to Einstein's equations, the two 4D boundaries feature tensions of the order of $\Lambda_5/k$, and a stabilisation mechanism is required \cite{Goldberger:1999uk} also to avoid a massless scalar mode, the radion.

First, we want to fix $m$ as required by the DD construction, so that \cite{Montero:2022prj}
\begin{equation}
    (\lambda m)^4 = \Lambda_4 = 10^{-122} M_\text{Pl}^4\,,
\end{equation}
where $\lambda$ is a generic coupling. Taking as a template the value $\lambda = 0.1$, we obtain
\begin{equation}
    m = 38~\text{meV}\;\; \text{for}\;\; \lambda = 0.1\,.
\end{equation}
The bulk cosmological constant and the 4D Planck mass are related as follows \cite{Randall:1999ee}:
\begin{eqnarray}
    \Lambda_5 &=& - 24 k^3 M_\text{Pl}^2\,, \label{eq:lambda5}\\
    M_\text{Pl}^2 &=& M_5^3/k \ (1-e^{-2 k r_c \pi})\,. \label{eq:M5}
\end{eqnarray}
For sizeable warping, i.e. $e^{-k r_c \pi} \ll 1$, the dependence on $r_c$ can be neglected and $M_5^3 = M_\text{Pl}^2 k$. Combining this relation with Eq.~\eqref{eq:lambda5}, we find:
\begin{equation}
    \Lambda_5 = - 24 \frac{M_5^9}{M_\text{Pl}^4}\,.
\end{equation}
\begin{figure}[tb]
\centering
\includegraphics[width=8cm]{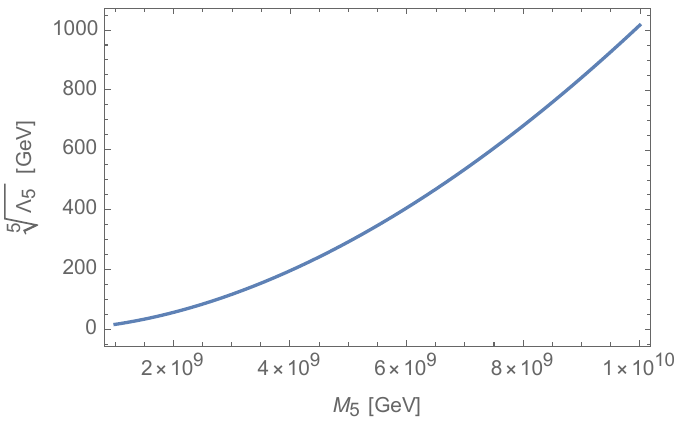}
\caption{\label{fig:lambda5} Scale associated to the 5D cosmological constant in the warped DD as a function of the fundamental Planck scale $M_5 \equiv \hat{M}$ within the range predicted by the Swampland.}
\end{figure}
If we identify $M_5$ with the species scale $\hat{M}$, then the scale associated to the 5D cosmological constant falls naturally in the ballpark of the electroweak scale,  generated by the Higgs mechanism, as shown in Fig.~\ref{fig:lambda5}. Interestingly, the warp factor $k$ is required to be a fraction of eV, hence falling in the same ballpark as the DD mass scale $m$. For instance, a benchmark
\begin{equation}
    \Lambda_5 = -(200~\text{GeV})^5\,,
\end{equation}
yields
\begin{equation}
    k = 450~\text{meV}\,. 
\end{equation}
The radius $r_c$ is fixed by the KK scale $m \sim k e^{- k r_c \pi} = 38$~meV, requiring 
\begin{equation}
    1/r_c = 570~\mbox{meV}
\end{equation}
and a moderate warp factor $k r_c = 0.78$. Notably, in the warped DD we consider, all parameters are of order a fraction of eV. Furthermore, a new value of the effective 5D Planck scale emerges:
\begin{equation} \label{eq:Mhat}
    M_5 = \hat{M} = 4.0 \times 10^{9}~\text{GeV}\,
\end{equation}
for the template values we considered. As already mentioned, consistency of the geometry in Eq.~\eqref{eq:metric5} with Einstein's equations requires the presence of brane tensions on the two boundaries of the extra dimension, with values
\begin{equation}
    |\Lambda_{\rm loc} | = |\Lambda_5|/k \sim \left(160~\mbox{TeV}\right)^4\,,
\end{equation}
for the template values. This value is consistent with the localisation of all SM fields on the boundaries, as it is much larger than the SM radiative contributions, which therefore does not destabilise the construction. It remains to be explained why the 5D cosmological constant value is much smaller than the fundamental Planck scale, $|\Lambda_5| \ll \hat{M}^5$.

Note that, for the emergence of the electroweak scale, a sizeable warping is essential. In fact, in the limit of flat space, i.e. $k r_c \pi \ll 1$, the 4D Planck scale and the KK mass would only depend on the radius $r_c$ \cite{Arkani-Hamed:1998jmv,Antoniadis:1998ig}:
\begin{equation}
    M^2_\text{Pl} \sim 2 \pi r_c M_5^3\,, \quad m \sim 1/r_c\,,
\end{equation}
while $\Lambda_5$ would be proportional to the small curvature $k$ as in Eq.~\eqref{eq:lambda5}. Hence, in this limit, the scale associated to the bulk cosmological constant would be much smaller than the hundreds of GeV.

\subsection{Doubly warped toy model}

Our strategy is now to apply the same Swampland arguments, based on the DC, to the 5D cosmological constant, which is much smaller than the fundamental Planck scale in the effective 5D theory described by the warped DD. Hence, we predict the presence of a second extra dimension characterised by a mass gap $m_5$, such that
\begin{equation}
   M_{\rm Pl}^{\alpha-5} |\Lambda_5| = (\lambda' m_5)^\alpha\,.
\end{equation}
In analogy with the DD, we will use $\alpha = 5$ in the following, even though this value violates the strong version of the DC in anti-de-Sitter space, which would require $\alpha \leq 2$ \cite{Lust:2019zwm}.  As a template, for $\alpha = 5$ we consider $\lambda' = 0.1$ and $m_5 = 2$~TeV, consistent with the previous estimates. Hence, within this context, discovering the KK modes at the TeV scale could be seen as an experimental test of the strong DC. If the strong DC must be satisfied, one additional dimension at the TeV scale may not suffice to obtain the correct value of $\Lambda_5$, hence it would be interesting to consider alternative constructions based on more than 6 dimensions, non-trivial geometries in the TeV dimensions, or direct contribution of string modes. We leave this exploration for future work.

In the $\alpha = 5$ case, we can again consider a Randall--Sundrum warped space, $\psi$, characterised by its own warp factor $\tilde{k}$ and radius $\tilde{r}_c$. The mass gap $m_5$ is related to the fundamental parameters of the 6D theory as follows:
\begin{equation}
    m_5 = \tilde{k} e^{- \tilde{k} \tilde{r}_c \pi}\,.
\end{equation} 
\red{For this warped geometry to remain a classical solution to Einstein's gravity, it is necessary that the curvature remains smaller than the quantum scale of gravity, which we identify with the species scale $\hat{M}$. Nevertheless, the two scales can be close enough to not introduce a new hierarchy}
If $\tilde{k} \lesssim \hat{M}$, then the 6D warp factor is required to be
\begin{equation}
    \tilde{k} \tilde{r}_c \lesssim 4.9\,.
\end{equation}
This moderate value means that it is possible to realize the TeV dimension with fundamental parameters all of order $\hat M$. For the same reason, the bulk cosmological constant in the 6D theory will also be of order $|\Lambda_6| \lesssim \hat{M}^6$, hence at its most natural value \red{close to the scale of quantum gravity effects}.

\subsection{Towards a realistic geometry}

The bottom-up approach we followed so far has led us to an interesting scenario with only two geometric fundamental scales: one emerging from the 4D cosmological constant, $m \sim \Lambda^{1/4} \sim 40$~meV, and another \red{close to the scale} corresponding to the fundamental Planck scale, $\hat{M} \sim 4 \times 10^9$~GeV. The electroweak scale, instead, emerges by the necessity to generate the 4D Planck scale $M_\text{Pl}$, which governs gravitational interactions at low energies and large scales. At the scale $\hat{M}$, the theory must consistently incorporate quantum gravity and it could correspond to a string theory scenario, compactified on a manifold with two larger space-like dimensions. Note that, although in our bottom-up approach we assumed warped geometry {\it \`a la} Randall--Sundrum, other warped geometries may also be consistent with our conclusions. Randall-Sundrum throats naturally  emerge in string compactification. Such constructions have played crucial roles in addressing issues like the moduli stabilisation problem, inflation, and the AdS/CFT correspondence. Attempts have also been made towards explaining the value of the 4D cosmological constant: for instance, two sub-millimetre extra dimensions have been considered \cite{Aghababaie:2003wz}, where a small cosmological constant emerges from 6D supergravity via self-tuning. Exact solutions for the 6D metric have also been sought for \cite{Carroll:2003db}, leading to a factorisable brane-world spacetime with two extra dimensions compactified on the two-sphere topology of a football. These approaches are substantially different from ours, where the two extra dimensions are asymmetric and the hierarchy in the compactification scales is justified by conjectures from Swampland arguments.

\begin{figure}[tb]
\centering
\includegraphics[width=8cm]{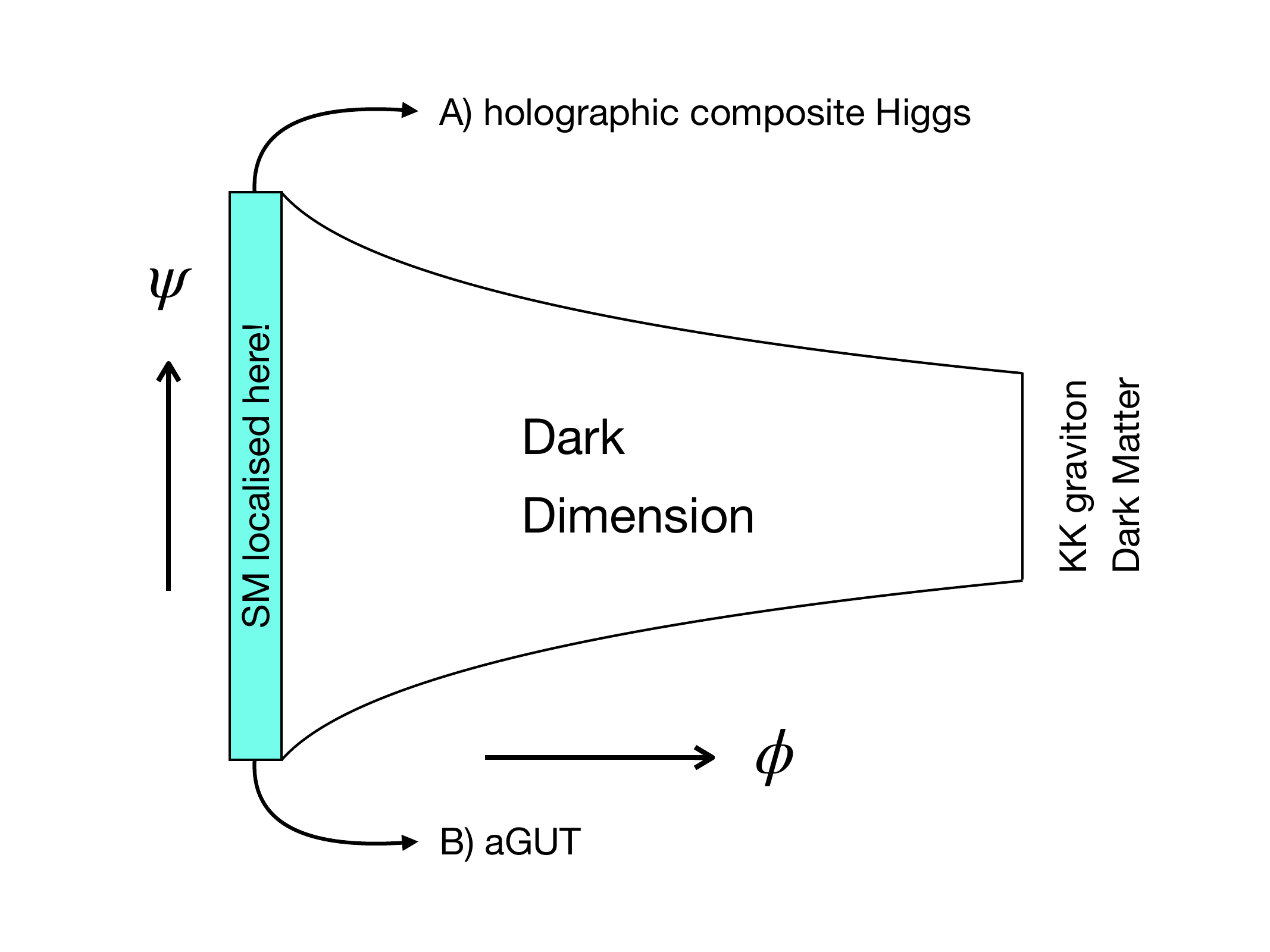}
\caption{\label{fig:6Dspace} Schematic illustration of the two extra dimensions predicted by the Swampland. The macroscopic DD corresponds to the horizontal coordinate $\phi$, while the TeV dimension is vertical $\psi$, shown not in scale. The usual 4D are perpendicular to the plane. The SM fields propagate on the $\phi=0$ 4D boundary and also probe $\psi\neq 0$, where holographic construction can be implemented, such as a composite Higgs or aGUT scenario.}
\end{figure}

We now provide a general perspective of the theory below the scale $\hat{M}$, \red{intended as the regime where quantum gravity effects are negligible}. The model consists of a six-dimensional space-time with a general warped background metric, such as 
\begin{equation} \label{eq:metric6}
    ds_6^2 = B(\psi) \left( A(\phi)\, \eta_{\mu \nu} dx^\mu dx^\nu + r_c^2 d\phi^2\right) + \tilde{r}_c^2 d\psi^2\,.
\end{equation}
The only requirement is that the warp factors $A$ and $B$ must be different, and provide hierarchical scales for the two extra dimensions, i.e. the meV scale for $\phi$ and the $10^{10}$~GeV scale for $\psi$. 
The analytic form of the metric functions $ A(\phi)$ and $B(\psi)$ is provided by the ultraviolet completion of the effective 6D model, such as string theory. Notice that the smallness of the 4D cosmological constant is responsible for the KK tower of light particles with mass $m$.
In the above 6D space, $\phi$ is the sub-eV DD with $k \sim 1/r_c \sim m$, and $\psi$ is the electroweak dimension with $\tilde{k} \sim 1/\tilde{r}_c \lesssim \hat{M}$. The SM fields, charged under gauge symmetries, are localised at $\phi =0$, hence living on a 5D brane characterised by a TeV scale warped extra dimension. Gravity and right-handed neutrinos, instead, probe the whole space, as illustrated in Fig.~\ref{fig:6Dspace}. 
\red{We remark that as long as $\tilde{k} \sim 1/\tilde{r}_c \lesssim \tilde{M}$ and quantum gravity effects are negligible, one could also find the metric from an effective 6D theory of gravity, as in the example below.}~\footnote{\red{A consistent metric~\eqref{eq:metric6} could also come as a solution of a complete theory of gravity, such as stemming directly for the compactification of string theory. Explicit examples are not known yet.}}

\vspace{0.5cm}

A 6D geometry consistent with Eq.~\eqref{eq:metric6} has been found by extending the original Randall-Sundrum model to  higher dimensions \cite{Choudhury:2006nj,Arun:2014dga,Arun:2015ubr}. The metric is obtained as solution of the Eintein's equations for a 6D action consisting of the Ricci scalar $R$ and a (negative) cosmological constant in the bulk: 
\begin{equation} \label{eq:Choudhury}
    S = \int {d^4 x} {d \psi} {d \phi}  \sqrt{-g_6} \big[ \hat{M}^4 R - \Lambda_6 \big] \,,
\end{equation}
plus terms localised on the 5D boundaries of the two-dimensional box.
The metric functions are 
\begin{equation}
    A(\phi) = e^{- 2 k\,r_c  |\phi|}\,, \;\; B(\psi) = \frac{\cosh^2\left(\tilde{k} \tilde{r}_c |\psi|\right)}{\cosh ^2\left(\tilde{k} \tilde{r}_c \pi \right)}\,,
\end{equation}
where \footnote{Note that we use dimension-full warp factors, while Ref.~\cite{Choudhury:2006nj} defines dimension-less ones. The matching between the two notation is done via the radii $r_c$ and $\tilde{r}_c$, respectively for $k$ and $\tilde{k}$.}
\begin{eqnarray}
\tilde{k}=\sqrt{\frac{-\Lambda_6}{10\, \hat{M}^4}}\,, \quad k=\frac{\tilde{k}}{\cosh{\tilde{k} \tilde{r_c} \pi}}\,.
\end{eqnarray}
While $\tilde{k}$ and $(\Lambda_6)^{1/6}$ are naturally close to the scale of $\hat{M}$, as needed, the warp factor $k$ is related to $\tilde{k}$ by the warping along the $\psi$ dimension. Hence, this solution implies $k \sim m_5 \sim 2$~TeV and it cannot describe our original scenario. If we wanted to adopt the geometry from \cite{Choudhury:2006nj}, one would need to raise the value of the fundamental Planck scale to $\hat{M} = (M^2_{\rm Pl} k)^{1/3} \sim 7 \times 10^{13}$~GeV and a much stronger warping in the DD, $k r_c \sim 10$ (and $\tilde{k} \tilde{r}_c \sim 8$). Our original scenario, therefore, motivates the search for double-warped geometries as solution from string theory, supported by the fact that stringy effects appear at the scale $\hat{M}$ and  cannot be neglected in the effective action Eq.~\eqref{eq:Choudhury}. We leave this exploration for future work.

\section{Enter the standard model}

In summary, our analysis points out a 6D scenario where all scales are of geometrical origin and they can be expressed in terms of two experimental inputs, the 4D cosmological constant $\Lambda$ and Planck scale $M_\text{Pl}$, and a few generic coupling constants. The DD mass scale and the fundamental Planck scale can be expressed as \cite{Montero:2022prj}
\begin{eqnarray}
    \hat{M} &=& \hat{\lambda}^{-1/3} \Lambda^{1/12} M_\text{Pl}^{2/3}\,, \\
    m &=& \lambda^{-1} \Lambda^{1/4}\,.
\end{eqnarray}
Furthermore, the electroweak scale is related to a 5D effective cosmological constant:
\begin{equation}
    m_5 = {\lambda'}^{-1} \sqrt[5]{\Lambda_5} \sim {\lambda'}^{-1} \lambda^{-3/5} \Lambda^{3/20} M_\text{Pl}^{2/5}\,,
\end{equation}
which has values naturally of the order of the TeV scale. The latter is a new result in a more general attempt to relate all physical scales in particle physics to the cosmological constant \cite{Gendler:2024gdo}.
To complete the picture, the metric parameters are given by
\begin{equation}
    k = \hat{\lambda}^{-1} \Lambda^{1/4}\,, \quad \tilde{k} \lesssim \hat{M}\,,
\end{equation}
while the radii are determined as:
\begin{equation}
    \pi k r_c = \ln \frac{k}{m}\,, \quad
    \pi \tilde{k} \tilde{r}_c = \ln \frac{\tilde{k}}{m_5} \sim \ln \frac{\hat{M}}{m_5}\,.
\end{equation}
The relations above fix all the relevant scales in the theory in terms of gravitational observable quantities.

To define a realistic model in the 6D space, we need to specify how the SM fields are defined. They propagate in a 5D manifold with warped geometry determined by $\tilde{k}$, characterised naturally by the TeV scale. As such, we foresee two possible realistic routes:
\begin{itemize}
    \item[A)] The 5D warped model may correspond to a holographic version of composite Higgs models \cite{Contino:2003ve,Agashe:2004rs,Hosotani:2005nz}. Such a model could also feature grand unification of the SM gauge couplings \cite{Agashe:2005vg}. For more details on phenomenological features we refer to Refs~\cite{Contino:2010rs,Panico:2015jxa,Cacciapaglia:2020kgq}.
    \item[B)] Another interesting possibility is offered by asymptotic Grand Unification (aGUT) scenarios in 5D \cite{Cacciapaglia:2020qky,Cacciapaglia:2023kyz}, which are compatible with a compactification scale around a few TeV thanks to baryon number conservation.  
\end{itemize}
In both cases, the vacuum expectation value of the Higgs boson is naturally tied to the mass scale of the microscopic extra dimension, $m_5$, hence solving the naturalness problem via geometrisation \cite{Randall:1999ee}. In case A, the Higgs emerges as a gauge-scalar \cite{Contino:2003ve}, while in B bulk scalars are also needed. The aGUT case has an additional attractive feature: due to the presence of ultraviolet fixed points for all bulk couplings \cite{Cacciapaglia:2023kyz}, the extra-dimensional field theory is under control up to $\hat{M}$ \cite{Gies:2003ic,Morris:2004mg}. 
Both scenarios also offer fertile ground for mechanisms of baryogenesis \cite{Bruggisser:2018mrt,Cacciapaglia:2020qky}, hence explaining the baryon asymmetry in the current Universe. Furthermore, potential sources of dark matter can be included. In holographic composite Higgs models, the main candidates are additional pseudo-Goldstone bosons, which can be produced thermally \cite{Frigerio:2012uc,Wu:2017iji,Balkin:2018tma} or asymmetrically \cite{Cai:2019cow}. In the aGUT case, massive modes are protected by baryon number and a relic density emerges during baryogenesis \cite{Cacciapaglia:2020qky}. Such dark matter components can complement the role of the bulk gravitons from the DD  \cite{Gonzalo:2022jac}.
In this paper, using the universal coupling of the SM brane to bulk gravitons, it is shown that spin-2 Kaluza-Klein modes are unavoidably produced during the cooling down of the Universe. The SM brane is initially in a thermal state after reheating, while bulk modes in the DD are unexcited. Then, the cooling of the SM brane leads to KK dark graviton production via gravitational couplings at a mass of about $1\div 50$~GeV, which subsequently decay to lighter ones with a mass about $1\div 100$~keV today. It has been shown that this mechanism can saturate the observed relic density measured in the cosmic microwave background. We should also stress that, as $m_5$ is of order TeV, both scenarios contains new particles with SM charges which could be observable at current (LHC) and future colliders. The specific signatures are model dependent, and they should be studied case by case.

\section{Conclusions}

In this paper we have shown that the Swampland's Dark Dimension, once realised in a warped background, leads to the emergence of the electroweak scale. The latter comes to life as an intermediate step between the cosmological constant scale, $10^{-2}$~eV, and a new fundamental Planck scale, $10^{10}$~GeV. This argument also requires a new extra dimension compactified at the TeV scale where standard model fields are allowed to propagate, hence providing new states that can be directly accessible at current (LHC) and future high energy colliders. Models of holographic composite Higgs or asymptotic Grand Unification can, therefore, be naturally realised alongside the dark dimension. In this scenario, the quantum contributions to the cosmological constant of both gravity and the SM of particle physics are tamed by Swampland arguments, i.e. the Distance Conjecture. Finally, a new unified and holistic scenario of fundamental interactions emerges from the Swampland, characterised by two fundamental scales. The electroweak scale and the Planck scale emerge by consistency. The two new scales are of geometric origin, hence their value is natural as long as the 6D geometry is appropriately stabilised. We recall that the main argument is based on consistency of the value of the cosmological constant, hence in this scenario gravity itself offers a new way to address the long-standing naturalness problem of the electroweak scale in the standard model of particle physics.

\vspace{0.2cm}

\section*{Acknowledgements}
We would like to thank theory division of CERN for hospitality. KBF would like to thank the Beijing Institute of Mathematical Sciences and Applications (BIMSA) for their hospitality, where part of this work was completed.

\bibliographystyle{utphys}
\bibliography{sample}

\end{document}